\documentclass[final,english]{bullsrsl}

\usepackage[latin1]{inputenc}

\usepackage[T1]{fontenc}

\usepackage{natbib} 

\usepackage{graphicx}
\usepackage{xcolor}

\begin{document}
\title{Photometric and Spectroscopic Study of Young Stellar Objects in IC~1848E Region}

\author[affil={1,2}, corresponding]{Shubham}{Yadav}
\author[affil={1}]{Neelam}{Panwar}
\author[affil={2}]{Lalita}{Sharma}
\author[affil={1,3}]{Rishi}{C}
\affiliation[1]{Aryabhatta Research Institute of Observational Sciences (ARIES), Manora Peak, Nainital, 263001, India}
\affiliation[2]{Department of Physics, Indian Institute of Technology Roorkee, Roorkee 247667, India}
\affiliation[3]{Mahatma Jyotiba Phule Rohilkhand University, Bareilly, 243006, U.P., India}

\correspondance{subham.yadav@aries.res.in}
\maketitle

\correspondance{your@email.address}


\maketitle

\begin{abstract}
We present a comprehensive analysis combining multi-epoch mid-infrared (MIR) photometric and optical spectroscopic observations of young stellar candidates associated with elongated, elephant trunk-like structures (ETLSs) 
in the IC~1848E star-forming region. A sample of 15 candidate YSOs, initially identified by 
\citet{2011PASJ...63..795C}, was investigated using multi-epoch NEOWISE $W1$ and $W2$ photometry 
to characterize their variability properties. After imposing stringent positional and photometric quality constraints, 
a final set of eight robust YSOs was retained and classified into three variability categories: non-variable, secular, and stochastic. One source, Star~\#~9, exhibits a pronounced curved variability in the MIR light-curve
 and color-magnitude domains, indicative of systematic evolution of its circumstellar disk 
and/or accretion processes. Optical spectroscopic data for this object show strong H$\alpha$ emission 
and prominent forbidden lines, providing clear evidence of active mass accretion and outflow phenomena. 
The combined photometric and spectroscopic diagnostics establish a close link between infrared variability, 
circumstellar disk evolution, and accretion-driven activity in YSOs associated with ETLSs in IC~1848E.
\end{abstract}





\section{Introduction}
Young stellar objects (YSOs) exhibit complex spectroscopic and photometric signatures that encode information about their evolutionary stage and ongoing accretion activity. These sources are intrinsically variable, with variability arising from a combination of accretion instabilities, magnetospheric processes, and structural changes within their circumstellar disks \citep{1994AJ....108.1906H, 2015ApJ...808...68H}.

Time-domain surveys in the mid-infrared (MIR) are particularly effective for probing such variability in embedded systems, where circumstellar dust reprocesses accretion-powered luminosity into longer wavelengths \citep{2011ApJ...733...50M, 2014AJ....147...82C}. 
The NEOWISE mission, with its long-term temporal coverage in the MIR, enables systematic characterization of both stochastic and secular variability in YSOs \citep{2019MNRAS.483.4424K, 2020MNRAS.499.1805L}. 
Variability-based classifications derived from these data provide important constraints on the physical mechanisms governing disk-star interactions across different evolutionary stages of YSOs \citep{2020MNRAS.495.3614C}.

In this study, we perform a comprehensive analysis of a sample of YSOs in the IC 1848E (W5 E) star-forming region (SFR), situated at an approximate distance of 2.1 kpc \citep{2011MNRAS.415.1202C,2011PASJ...63..795C}. This H{\sc ii} region harbors a rich population of young stars \citep[][ hereafter K08]{2008ApJ...688.1142K} and contains a substantial number of bright-rimmed clouds (BRCs) as well as elephant trunk-like structures (ETLSs). The morphology of the ETLSs, combined with their preferential alignment toward the ionizing O-type star, provides strong evidence that their formation and subsequent evolution are closely governed by radiative feedback processes \citep{2011PASJ...63..795C,2019PASJ...71..121S}.
One or two sources are commonly associated with the tips of these ETLSs, which are believed to form as a consequence of hydrodynamical instabilities that develop at advancing ionization fronts \citep{2011PASJ...63..795C,2012A&A...546A..74D,2012MNRAS.426..203H}. Although photometric and spectral energy distribution analyses indicate that these ETLS sources are young, low-mass stars; detailed time-series photometric and spectroscopic observations are required to obtain direct diagnostics of their pre-main-sequence (PMS) evolutionary phase and to characterize the accretion processes currently operating in these systems.

In this work, we combine optical spectroscopy with NEOWISE MIR photometry to characterize the accretion activity and photometric variability of the ETLS-associated sources in the IC~1848E. Target selection and observations are discussed in Section 2. In Section 3, we discuss the results of our NEOWISE and Spectroscopic data. Finally, in Section 4, we present the summary and conclusion of the present work.
\section{Target Selection, Observations and Data Used}
\subsection{Target Selection}
ETLSs are morphologically distinct from classical bright-rimmed clouds in terms of their extreme aspect ratios and narrow widths \citep[see,][] {1991IAUS..147..498S,1994ApJS...92..163S,2011PASJ...63..795C}. These ETLSs were identified in Spitzer images through a visual inspection aimed at detecting elongated structures that host one or more stellar sources at their tips. In this work, we studied 15 sources associated with ETLSs in the IC 1848E region, as originally reported by \citet[][]{2011PASJ...63..795C}. Fig. 1(a) presents the spatial distribution of these sources, which are marked with green circles.

\begin{figure*}[!t]
\centering
\includegraphics[width=\linewidth]{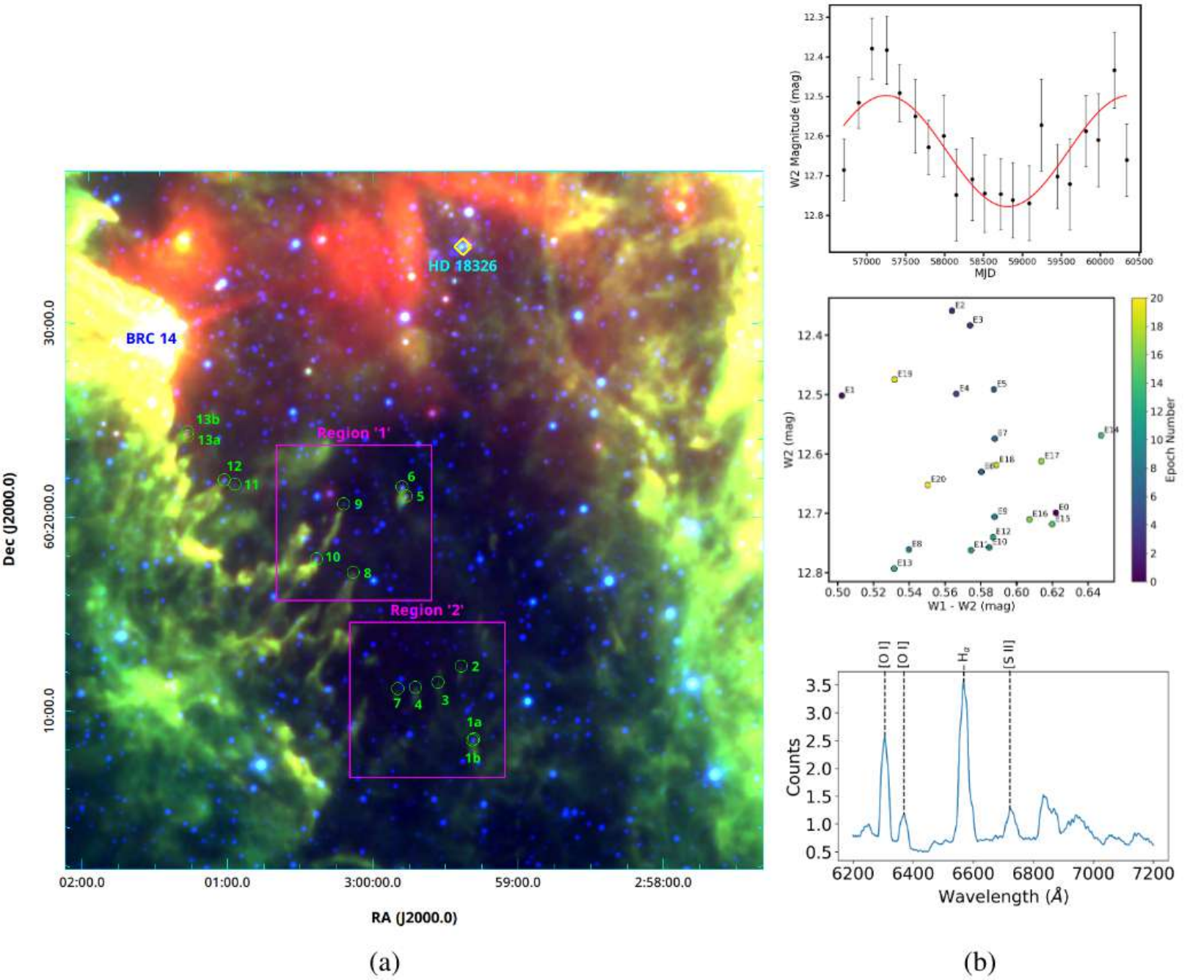}
\caption{(a) WISE three-color composite image (red: 22 $\mu$m, green: 12 $\mu$m, blue: 4.6 $\mu$m) of the region. The positions of the stars located at the tips of the ETLS, as identified by \citet[][]{2011PASJ...63..795C}, are indicated by green circles. (b) Multi-epoch MIR variability (top panel), color-magnitude evolution (middle panel) and optical spectra (bottom panel) of Star~\#~9 in the IC~1848E region.}
\label{fig:main_layout}

\end{figure*}

\subsection{MIR Photometric Data}
The NEOWISE mission provides multi-epoch MIR photometry in the WISE W1 (3.4~$\mu$m) and W2 (4.6~$\mu$m) bands, making it particularly well suited for probing variability in YSOs embedded in dusty environments. Its recurrent observations at roughly six-month intervals since late 2013 \citep{2014ApJ...792...30M} provide a basis for probing long-term photometric variability associated with changes in the accretion process and the evolution of the circumstellar disk.
Furthermore, each observing epoch consists of multiple exposures obtained over a short timescale, allowing robust construction of MIR light-curves while minimizing the impact of short-term photometric fluctuations. We downloaded the NEOWISE data for our targets from the NASA/IPAC InfraRed Science Archive.
\subsection{Optical Spectroscopic Data}
Optical spectroscopy of the selected ETLS-associated stars was carried out using the low-resolution DOLORES spectrograph \citep{2014SPIE.9147E..4XG} mounted on the 3.6-m Telescopio Nazionale Galileo, employing multi-object spectroscopy with the LR-R grism during the year 2011. This configuration provides a spectral resolution of $\lambda/\Delta\lambda \sim 700$ and a wavelength coverage spanning the range 4500--10000~\AA\ .Spectroscopic data were acquired for targets in two multi-object spectrograph fields, with total integration times of 2400s and 3600s, respectively. Contemporaneous arc-lamp exposures were also acquired. The data reduction and wavelength calibration procedures were performed using the IRAF software package.

\section{Results}
\subsection{Photometric Results}

We used NEOWISE MIR photometry to investigate variability in sources associated with IC 1848E by analyzing multi-epoch W1 and W2 light curves of 15 ETLS stars. For reliable variability analysis, only sources detected in more than five W2 epochs were considered. We further applied strict positional constraints, retaining only NEOWISE exposures within $2\sigma$ of the mean positional offset and with a positional standard deviation below $0.3^{\prime\prime}$. Seven sources failed these criteria and were excluded, leaving eight candidates for light-curve analysis.

YSO variability was classified into two broad categories, secular and stochastic, based on light-curve morphology. Secular variables exhibit long-term trends or periodic behavior and were further divided into linear, curved, and periodic subclasses. Stochastic variables lack regular temporal patterns and were classified as burst, drop, or irregular, representing short-lived brightening, short-lived fading, and random non-periodic variations, respectively.

Based on their MIR light curves, five sources show no significant brightness variations and were classified as non-variable over the NEOWISE observing baseline. Of the remaining sources, one exhibits linear secular variability, one shows a clear curved trend, and one displays irregular variability without a well-defined long-term pattern. Figure~1b (top panel) presents the NEOWISE light curve of Star~\#9, classified as a curved-type variable, while the middle panel shows its temporal evolution in both color and brightness.

\subsection{Spectroscopic Results}
We analyzed the optical spectra of five YSOs associated with the IC~1848E region using standard spectroscopic reduction procedures. All sources exhibit clear H$\alpha$ emission, confirming their PMS nature and ongoing accretion activity. The equivalent width (EW) of H$\alpha$ emission line was measured for all YSOs to quantify the strength of accretion-related emission. As a representative case, the optical spectrum of Star~\#~9 is presented in Fig. 1(b) (bottom panel). This source exhibits an H$\alpha$ emission-line EW of approximately $60$~\AA. Given that the EW of chromospheric H$\alpha$ emission is strongly dependent on spectral type, increasing from $\sim$3.9~\AA\ for early-K stars to $\sim$41.9~\AA\ for M7 stars \citep{2003AJ....126.2997B}, the measured EW of Star~\#~9 exceeds the expected chromospheric contribution over the entire K--M spectral range. This implies that its H$\alpha$ emission is predominantly powered by magnetospheric accretion, thereby supporting its classification as a classical T-Tauri star (CTTS).

\begin{table}
\centering
\small
\begin{tabular}{ccc|ccc}
\hline
Star ID & H$\alpha$ emission  & NEOWISE variability&Star ID & H$\alpha$ emission& NEOWISE variability\\
       & in optical spectra &                      &        & in optical spectra &\\
\hline
1b & ..., Class III (K08)  & Non-Variable & 10 & Yes, CTT features & Non-Variable\\
5 & Yes, CTT features  & Linear&11 & Yes, CTT features & Non-Variable\\
6 & Yes, CTT features  & Non-Variable&12 & ..., Class II (K08) & Irregular\\
9 & Yes, CTT features  & Curve&13b & ..., Class II (K08) & Non-Variable \\
\hline
\end{tabular}
\caption{Optical spectroscopic and MIR variability features of the sources having NEOWISE data.}
\end{table}

The remaining YSOs also show H$\alpha$ emission with EW values exceeding 10~\AA, characteristic of actively accreting CTTS. The widespread presence of strong H$\alpha$ emission indicates sustained mass accretion from the circumstellar discs. In addition, forbidden [O~I] emission at 6300~\AA\ and 6363~\AA\ is detected in several sources, particularly those with stronger accretion signatures, while the detection of [S~II] emission provides further evidence of outflow activity. A detailed spectroscopic analysis of the remaining sources is in preparation (Yadav et al. 2026). Overall, the spectroscopic diagnostics indicate that the stellar sources in IC~1848E are in an early evolutionary stage, undergoing active mass accretion and outflow processes.

\section{Summary and Conclusion}
We investigated the variability and spectroscopic signatures of sources associated with ETLSs in the IC~1848E using MIR photometry and optical spectroscopy. Of the 15 sources identified by \citet{2011PASJ...63..795C}, 10 met the NEOWISE positional quality criteria. After applying the criterion of at least 5 epochs of data, we were left with 8 sources. The MIR light curves reveal diverse variability, including non-variable, secular, stochastic, and mixed behavior, reflecting a range of accretion- and disk-related processes. In particular, Star~\#~9 exhibits a well-defined curved-type variability, accompanied by correlated changes in infrared color (W1$-$W2) and brightness (W2) in its multi-epoch color--magnitude diagram (CMD). Such behavior is consistent with variations in accretion and/or the inner circumstellar disk \citep{2015AJ....150..118P}. Although the CMD alone cannot uniquely identify the underlying mechanism, together with the spectroscopic evidence, it supports time-dependent changes in the inner circumstellar environment.

Optical spectroscopic observations confirm the young nature of the YSOs through strong H$\alpha$ emission. The detection of forbidden [O~I] and [S~II] emission lines provides additional evidence for disk winds and mass-loss activity, supporting the presence of active disk--accretion processes inferred from the photometric variability. Taken together, the spectroscopic diagnostics and MIR photometric variability provide mutually consistent evidence that the observed variability is linked to ongoing accretion-related activity and the evolution of the inner circumstellar environment.
Overall, our findings indicate that ETLS-associated YSOs in IC~1848E are dynamically evolving systems whose observed variability and spectral properties are shaped by time-dependent disk-accretion physics under the influence of radiative feedback.
\begin{acknowledgments}
We express our sincere gratitude to the anonymous referee for constructive comments and suggestions. SY thanks Mr. Tarak Chand for his valuable help and suggestions in spectroscopic data reduction. NP expresses profound gratitude to Prof. K. Ogura, the late Dr. A. K. Pandey, and Dr. Luca Di Fabrizio for their encouragement and substantial support during the TNG observations. NP and RC also acknowledge financial support from the Science and Engineering Research Board, Government of India, under grant number CRG/2021/005876. This publication makes use of data products from the Near-Earth Object Wide-field Infrared Survey Explorer (NEOWISE), which is a joint project of the Jet Propulsion Laboratory/California Institute of Technology and the University of Arizona. This work was presented at the BINA-2025 workshop.
\end{acknowledgments}
\begin{furtherinformation}
\begin{orcids}
  \orcid{0000-0002-0151-2361}{Panwar}{Neelam}
  \orcid{0009-0007-6397-3220}{Rishi}{C.}
\end{orcids}

\begin{authorcontributions}
All authors have made substantial and meaningful contributions to the conception, execution, and/or interpretation of this work.
\end{authorcontributions}
\begin{conflictsofinterest}
The authors declare no conflict of interest.
\end{conflictsofinterest}
\end{furtherinformation}

\bibliographystyle{bullsrsl-en.bst}

\bibliography{extra.bib}

\end{document}